\documentclass[conference,10pt]{IEEEtran}
\IEEEoverridecommandlockouts

\usepackage{graphicx}
\usepackage{epsfig}
\usepackage{color}
\usepackage{caption}
\usepackage{subcaption}
\usepackage{algorithm}
\usepackage{algorithmic}
\usepackage{amssymb}
\usepackage{amsmath}
\usepackage{epstopdf}
\usepackage{latexsym}

\usepackage{cite}
\usepackage{amsmath,amssymb,amsfonts}
\usepackage{algorithmic}
\usepackage{graphicx}
\usepackage{textcomp}
\usepackage{xcolor}

\makeatletter
\newcommand\fs@spaceruled{\def\@fs@cfont{\bfseries}\let\@fs@capt\floatc@ruled
  \def\@fs@pre{\vspace{5\baselineskip}\hrule height.8pt depth0pt \kern2pt}%
  \def\@fs@post{\kern2pt\hrule\relax}%
  \def\@fs@mid{\kern2pt\hrule\kern2pt}%
  \let\@fs@iftopcapt\iftrue}
\makeatother

\begin{document}



\title{\huge AI-enabled Blockchain: An Outlier-aware Consensus Protocol for Blockchain-based IoT Networks}



\author{\IEEEauthorblockN{Mehrdad Salimitari, Mohsen Joneidi, and Mainak Chatterjee
\IEEEauthorblockA{{Department of Electrical Engineering and Computer Science}\\
{University of Central Florida}, Orlando, FL 32816\\
Email: \{mehrdad@cs, joneidi@ece, mainak@cs\}.ucf.edu}
\thanks{This research was partially funded by Cyber Florida's Collaborative Seed Award program.}
\vspace{-5mm}
}
}

 \maketitle

\begin{abstract}
A new framework for a secure and robust consensus in blockchain-based IoT networks is proposed using machine learning. Hyperledger fabric, which is a blockchain platform developed as part of the Hyperledger project, though looks very apt for IoT applications, 
has comparatively low tolerance for malicious activities in an untrustworthy environment. 
To that end, we propose AI-enabled blockchain (AIBC) with a 2-step consensus protocol that uses an outlier detection algorithm for consensus in an IoT network implemented on hyperledger fabric platform.
The outlier-aware consensus protocol exploits a supervised machine learning algorithm which detects anomaly activities via a learned detector in the first step. Then, the data goes through the inherent Practical Byzantine Fault Tolerance (PBFT) consensus protocol in the hyperledger fabric for ledger update.
We measure and report the performance of our framework with respect to the various delay components. 
Results reveal that our implemented AIBC network (2-step consensus protocol) improves hyperledger fabric performance in terms of fault tolerance by marginally  compromising the delay performance.



\end{abstract}


\vspace{-2mm}

\section{Introduction}
\label{sec-introduction}

The wide spread deployment of IoT devices is enabling the automation of various aspects in our daily lives. 
One of the success stories of IoTs has been the automation of homes and cities which are referred as \textit{smart homes} and \textit{smart cities}. 
A significant obstacle towards realizing the true vision of smart homes is securing the 
communicated and stored data in the home network against malicious acts desiring to wreak havoc with someone's home~\cite{fernandez2018review}. 


Though there are many competing technologies that try to immune data in smart homes against attacks, 
blockchain has emerged as probably the most promising for both 
i) securing the home network against manipulation attacks on stored data and
ii) providing a secure platform for all the devices in the network to communicate with each other~\cite{lin2017using, novo2018blockchain}. 
In a blockchain, the data is immutable because of the underlying consensus protocols-- a process by which all transactions are validated by all the nodes~\cite{dai2019blockchain}.
Therefore, manipulation attacks on transmitted or stored data are not plausible through a single compromised node and majority of nodes ought to be compromised for a successful attack~\cite{salimitari2017profit}.
Different consensus protocols and their applicability towards
IoT networks can be found in~\cite{salimitari2018overview,fernandez2018review}.


Using blockchain for the IoT devices in a smart home is non-trivial. 
This is primarily because the IoT devices are 
 resource-constrained and might not be able to perform the extensive 
 computations required to achieve consensus. For example, 
 blockchain consensus protocols such as Proof of Work is contingent upon solving compute-intensive hash functions, having storage requirements, and needing low-latency communication links~\cite{reyna2018blockchain}.
 Resource-constrained devices cannot fulfill these requirements~\cite{zheng2016blockchain}. Recently, several novel consensus approaches have been proposed to overcome these limitations. One of the projects which tries to address these challenges is the 
 hyperledger project~\cite{salimitari2018overview}--
 which is what we chose as our evaluation platform in this research.

We implement a blockchain-based IoT smart home network on
hyperledger fabric which has low computational requirements and fast network response time that makes it desirable for IoT applications~\cite{salimitari2018overview,zheng2016blockchain}. 
Hyperledger fabric uses practical byzantine fault tolerance (PBFT) method which can successfully reach consensus over new data if the ratio of malicious devices is less than $1/3$~\cite{zheng2017overview}. On the other hand, most of the devices in a home network are resource constrained and very vulnerable to different types of cyber-attacks. Therefore, we should define a mechanism to detect malicious activities and the compromised devices and disconnect them from  the rest of the network in order to exclude them from participating in the PBFT consensus protocol. 

\begin{figure}[!h]
\centering
\includegraphics[width=3.0in, height=1.7in]{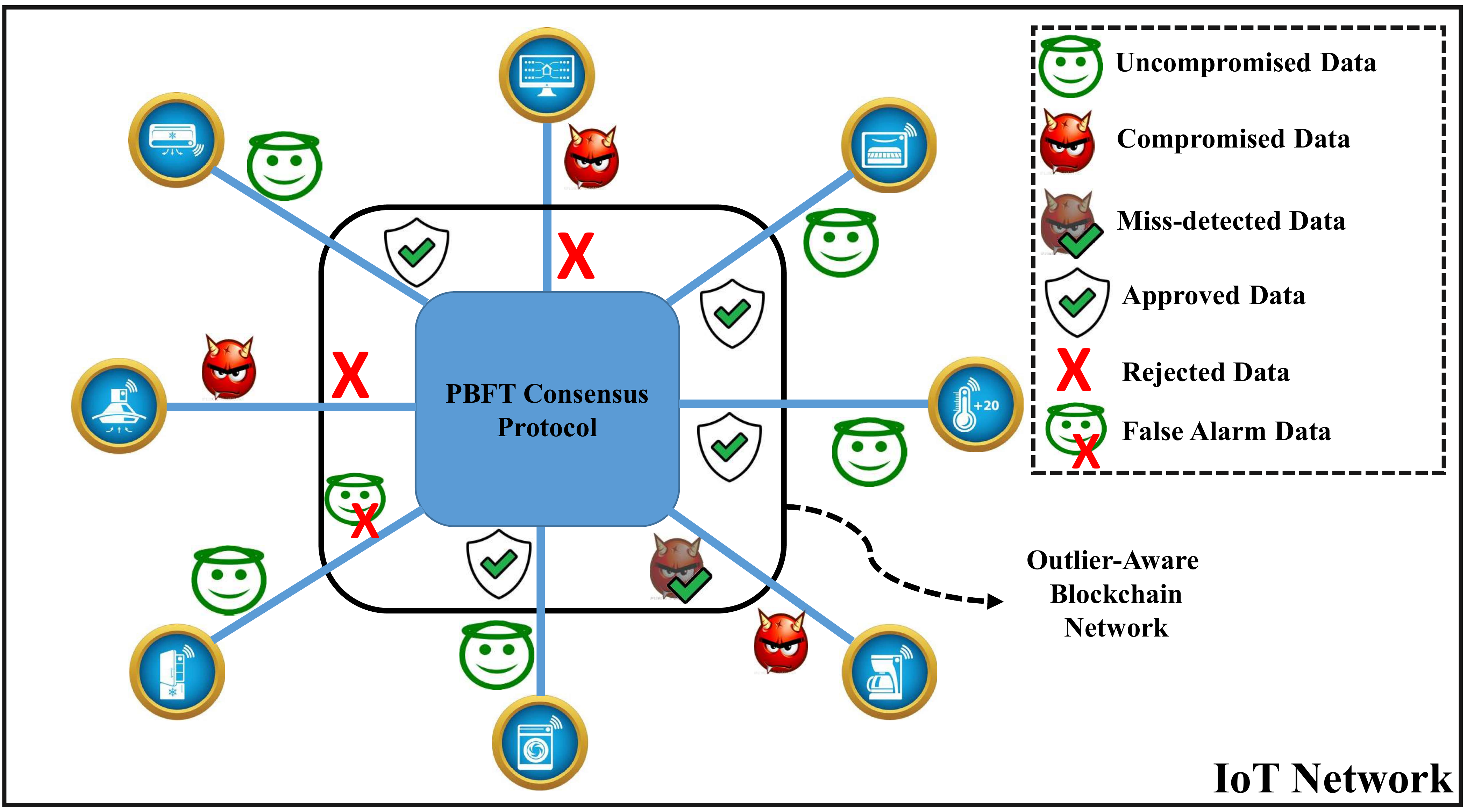}
  \caption{\small{The 2-step consensus protocol in our proposed AIBC network. The attack ratio in this schematic is $37.5\%$ which PBFT cannot tolerate in a conventional blockchain network. However, the AIBC network with outlier detector will attenuate the attack impact to $20\%$ in the first step which can successfully go through PBFT as the second step consensus protocol.}}
\label{maintopology}
\vspace{-4mm}
\end{figure}

In this paper, we propose the AI-enabled blockchain (AIBC) network with a 2-step consensus protocol using an outlier detection algorithm as the first step and PBFT as the second one.
Outlier detection algorithm discovers anomalies in multimodal data that is captured by the various IoT devices in a smart home network.
We use \textit{multimodal data fusion} to map the different data types received from different devices to an intermediate domain.
Since the captured data from different sensors are not independent of each other, the outlier detection algorithm aims to learn the intrinsic structure of the data and exploits the inter-dependencies among data from different devices. 
Eventually, outlier data is rejected at the end of the first step and the corresponding device is refrained from going to the second consensus step (PBFT). Thus, the overall fault tolerance of the blockchain network is enhanced in comparison to a naive hyperledger fabric implementation. Fig.~\ref{maintopology} shows the intuition of our proposed 2-step consensus protocol.

To the best of our knowledge, this is the first research work that tries to practically apply machine learning to reach consensus in a blockchain network. Gupta et al. have suggested the possibility of applying machine learning techniques to blockchain's consensus process as a future research work without any investigation~\cite{gupta2019blockchain}. Dinh et al. have discussed the benefits of integration of blockchain and artificial intelligence (AI)~\cite{dinh2018ai}. They suggest that a blockchain governed by a machine learning algorithm might be able to detect attacks and invoke proper defence mechanisms or isolate the compromised component. We have successfully formulated and implemented this idea called AI-enabled blockchain (AIBC). Dey has proposed a utility function similar to the function used in~\cite{salimitari2017profit} to detect anomaly~\cite{dey2018securing}. Then, he claims this value can be used as a feed for a supervised machine learning algorithm to measure the likeliness of an attack and prevent the blockchain confirmation of that transaction in the consensus protocol. However, he does not propose an algorithm or implementation to design a practical consensus protocol.

In order to test the validity of our 2-step consensus protocol for IoT networks, we implement a 3-layer architecture using hyperledger fabric. The first layer is the application layer containing different IoT devices. The second and third layers are the edge blockchain layer and the core blockchain layer
that contain  different components of the AIBC. 

We measure the latency of different parts of our implementation and the number of outlier devices in each time unit using synthetic data represented as  a  matrix. We compare our proposed architecture with the conventional hyperledger fabric implementation and investigate the effect of the outlier detection algorithm on fault tolerance.
Results reveal that our proposed AIBC network can reach consensus over new data in milliseconds with better fault tolerance than a naive hyperledger fabric implementation. This is achieved by detecting the malicious devices in the first consensus step and prohibiting them from participating in the PBFT step.

\section{Proposed System Architecture}
\label{sec-topology}

In this section, we discuss the proposed 3-layer architecture and the communication protocol between the different components of the implemented AIBC network. 


\subsection{Three-layer Architecture}
\begin{figure}[!b]
\centering
\vspace{-3mm}
\includegraphics[width=3.4in, height=1.8in]{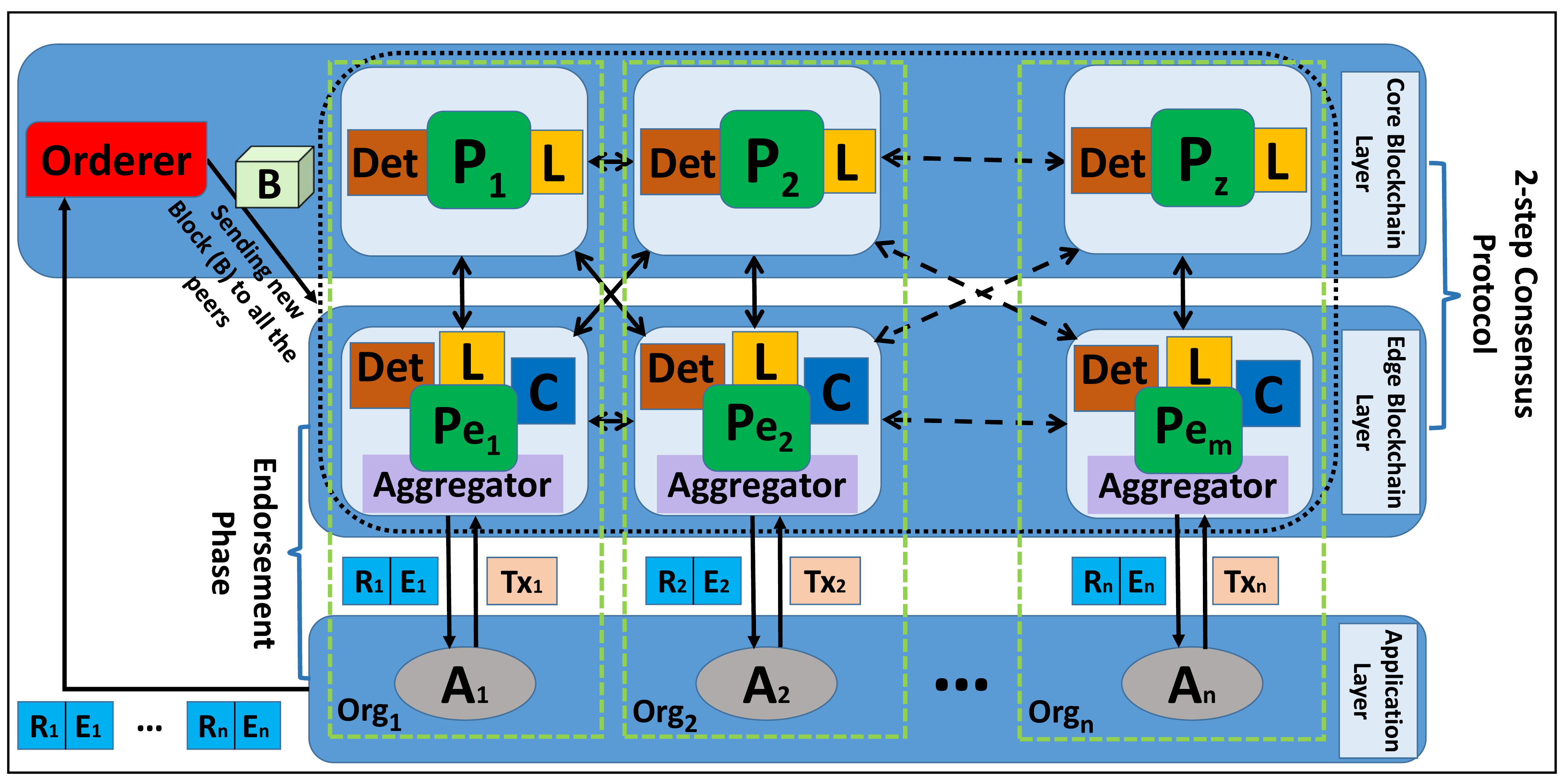}
   \caption{\small{The topology of our implemented 3-layer AIBC network.}}
\label{layers}
\vspace{-4mm}
\end{figure}

The topology of our implemented 3-layer network is illustrated in Fig.~\ref{layers}. The first layer is the application layer containing $n$ smart devices, smart meters, and sensors ({\bf$A_1$} to {\bf$A_n$}) within a smart home network. The second layer is the edge blockchain layer that includes $m$ endorsing peers ({\bf$P_{e_1}$} to {\bf$P_{e_m}$}) and data aggregators. 
This layer endorses the new transactions ({\bf$Tx_1$} to {\bf$Tx_n$}) from applications
and is partially responsible for the  2-step consensus protocol.
The third layer is the core blockchain layer consisting of an orderer and $z$ regular peers ({\bf$P_1$} to {\bf$P_z$}). 
This layer creates a block out of the received endorsed transactions ({\bf$R_1/E_1$} to {\bf$R_n/E_n$}, where $R$ and $E$ denote transaction results and their corresponding endorsements respectively) from the application layer and is also responsible for the 2-step consensus protocol.

There are $n$ organizations ($Org_1$ to $Org_n$) in the AIBC network each of which contains one application, at least one endorsing peer ($n \le m$), none or few regular peers ($n \le z$ or $n \ge z$ ). For ease of illustration, only one regular peer and one  endorsing peer are shown in each organization in Fig.~\ref{layers}. Endorsing peers have an aggregator to receive data (transactions) from the application layer,  a copy of the chaincode (also known as smart contract, shown as {\bf C}) to endorse new transactions, a copy of the ledger (shown as {\bf L}), and the detector (shown as {\bf Det}) to participate in the 2-step
consensus protocol. Regular peers have only one copy of the ledger and the detector for participation in the 2-step consensus protocol to enhance the security of the blockchain network.
Peers (both endorsing and regular) are involved  in the 2-step consensus protocol. Thereby, as the number of peers ($m$ and $z$) increases, more of them ought to be infected in order to prevent a successful consensus.  

\subsection{Communication protocol in AIBC network}
There exist two different communication protocols in the AIBC network: {\it query process} and {\it invoke process}. 

\subsubsection{Query Process}
In query process, an application connects to an arbitrary endorsing or regular peer to get updated about the current state of the ledger. This process requires only three steps: 
(i) the application connects to a peer directly (without an aggregator), 
(ii) the application sends query request to the peer, and 
(iii) the peer accesses its copy of ledger and sends the result back to the application.

\subsubsection{Invoke Process}
\begin{figure}[t]
\centering
\includegraphics[width=3.5in,height=2 in]{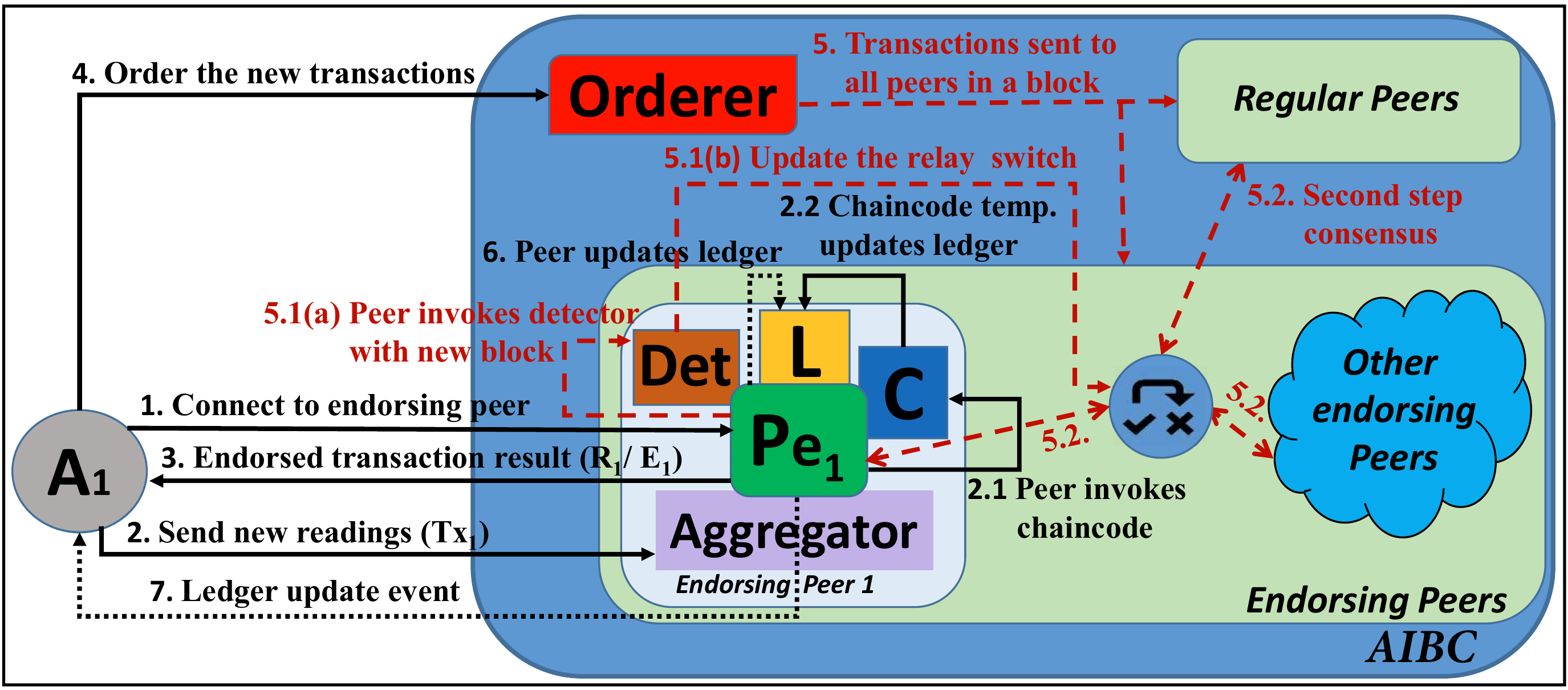}
   \caption{\small{The invoke process communication protocol in the implemented AIBC network using a relay switch.}}
\label{topology}
\vspace{-5mm}
\end{figure}

During this process, an application connects to its corresponding endorsing peers to request a change in the ledger by its new data. A simplified version of our implementation with just one application ({\bf $A_1$}) and its corresponding endorsing peer ({\bf$P_{e_1}$}) is delineated in Fig.~\ref{topology} to clarify the communication protocol in an invoke process. It should be noted that there are other applications, other endorsing peers and regular peers as shown in Fig.~\ref{layers}. Other endorsing peers communicate with their corresponding application and simultaneously traverse all the communication steps shown in Fig.~\ref{topology} (steps 1 to 7) whereas the regular peers go through steps 5 and 6 only.
The invoke process has three phases: 
(i) the endorsement phase (denoted with solid lines),
(ii) the 2-step consensus protocol (denoted with dashed lines) and
(iii) the ledger update (denoted with dotted lines) in Fig.~\ref{topology}.

\vspace*{5pt}
\noindent
{\bf \underline{Endorsement Phase:}}
This phase has 4 steps:
\\
{\bf 1.} In this step, each application needs to connect to one or more endorsing peers according to the endorsement policy (not arbitrarily). Endorsement policy delineates which endorsing peers from which organizations are required to endorse a new transaction ({\bf $Tx$}) proposed by an application ({\bf $A$}). In our implementation, each application communicates with only one endorsing peer as shown in Fig.~\ref{layers}.\\
{\bf 2.} The application ({\bf$A_1$}) reads from a smart device, and sends its new readings to its corresponding endorsing peer (shown as {\bf $P_{e_1}$}) in the AIBC network.

{\bf 2.1.} The endorsing peer executes the new transaction (data) in its copy of chaincode to endorse it.

{\bf 2.2.} If the transaction get endorsed successfully, the chaincode temporarily updates that endorsing peer's copy of ledger as a proposed update.
\\
{\bf 3.} The endorsing peer sends the result of the proposed transaction ({\bf$R_1$} ) along with its endorsement ({\bf$E_1$} ) obtained by the endorsing peer's digital signature to the corresponding application.
\\
{\bf 4.} The application checks all the received results along with their corresponding endorsements from corresponding endorsing peers to have the same result with valid endorsements. If these requirements are met, the application sends the result with its corresponding endorsements to the orderer to be added in the new block.

\vspace*{5pt}
\noindent
{\bf \underline{2-step Consensus Protocol}:} The two steps of the consensus protocol consist of the detector 
(which detects the outliers) and execution of PBFT. 

\noindent
{\bf 5.} The orderer combines all the received transaction results and their corresponding endorsements in a new block and sends it to all the peers (both endorsing and regular) to initiate the 2-step consensus protocol. 

\noindent
{\em Step 1:}  {\em Detector.}\\
{\bf 5.1(a)} Each peer checks all the transactions within the block via its \textit{detector} that uses the outlier detection algorithm discussed in Section~\ref{algorithm-sec} to find outlier data and reject it.
\\
{\bf 5.1(b)} The detector excludes the peers associated with the organization containing the application that generated the outlier data from participating in the second step consensus. To do so, 
the detector updates 
the relay switch in order to notify its corresponding peer which 
set of peers (both endorsing and regular)  it must connect to for the
second step consensus. This step can prevent {\em more than $33.\overline{3}\%$}
of compromised nodes to intervene in the PBFT consensus protocol
to some extent-- the exact fraction depends on the  accuracy of the detector
(discussed in Section~\ref{sec-performance}). 

\noindent
{\em Step 2:} {\em PBFT.}\\
{\bf 5.2} Each peer verifies whether each of the transactions in the new block is endorsed by all the required peers specified in the endorsement policy. In addition, they check if the result of a specific transaction is the same from all the required endorsing peers. This is to ensure that the application is not compromised and has not sent an incorrect result for the transaction. All the transactions in the block are labeled as valid or invalid after verification of the endorsements by each of the peers. Then, the peers connects to their trusted set of peers according to the obtained relay switch in the previous step to reach a consensus over the new block using PBFT method.

\vspace*{5pt}
\noindent
{\bf \underline{Ledger Update}: }After the completion of the consensus protocol, the invoke process is
finalized as:\\
{\bf 6.} Each peer updates its copy of ledger.\\
{\bf 7.} Applications get notified about the ledger update.

\section{Outlier Detection}
\label{algorithm-sec}
In order to detect malicious devices, we propose to employ machine learning for  outlier detection  as the first step of consensus. To that aim, the low-rank assumption is considered as the core model which is a popular assumption in machine learning \cite{udell2016generalized}. We implement the outlier detector using a secondary chaincode which is installed on all the peers in the blockchain network (shown by Det in Fig.~\ref{layers}). To apply the outlier detection algorithm to the multi-modal data gathered from different sensors and devices, we need to first fuse the data for a unified representation.

\subsection{Multimodal Data Fusion}
Sensors and devices in a smart home deal with different types of data.  To impose a model for the ensemble of data in AI-enabled systems, we need to map the data to a meaningful intermediate domain~\cite{baltruvsaitis2018multimodal}. 
The data from device $n$ at time slot $t$ is denoted by $\boldsymbol{d}^n_t \in \mathbb{R}^{b_n}$ where ${b_n}$ is the dimension of received data from sensor $n$. Let $\boldsymbol{d}_t \in \mathbb{R}^{b}$ represent the fused data  at the $t^{\text{th}}$ time slot where $b=\sum b_n$. Matrix $\boldsymbol{D}\in \mathbb{R}^{b\times T}$ represents collected data from $N$ devices over $T$ time slots. In other words, $\boldsymbol{D}=[\boldsymbol{d}_1 \cdots,\;\boldsymbol{d}_T$].   

\subsection{Outlier Detection Algorithm}
For detecting inconsistent data, we use low-rank data structure which is a typical assumption for many real-life data~\cite{joneidi2016union}. This model is one of the most well-known data structures in signal processing and data mining~\cite{esmaeili2019novel}. However, considering the recent advances in AI, deep learning methods and non-linear models can also  be imposed~\cite{li2018multi,zhao2018consensus}. Without loss of generality, the low-rank model is assumed in the present paper.
\begin{table}[b]
\vspace{-3mm}
\caption{\small{Summarization of  parameters in the employed model.}}
\centering
\small
\begin{tabular}{ |p{1.55cm}|p{1.75cm}|p{4cm}|}
 \hline
\bf{ Model's parameter} & \centering {\vspace{0.1pt}  \bf{Variable}} &   \vspace{0.1pt} \qquad \qquad \bf{Description}\\
 \hline
 \centering
 {\vspace{1.6mm} \bf{Span}}   & \vspace{1.6mm} $\boldsymbol{U}\in \mathbb{R}^{b\times R}$    & Span of the model (regular patterns are linear combination of these bases.)\\
 \hline
 \centering
\bf{Estimated rank}   & \vspace{0.3mm} \qquad $R_{est}$  & The number of basic patterns in the span.\\
 \hline
 \centering
{\vspace{1.6mm}\bf{Threshold}}    & \vspace{1.6mm} \quad $\boldsymbol{h}\in \mathbb{R}^{b}$  & Threshold for the margin between trustworthy and outlier data. \vspace{+1mm}\\
 \hline 
\end{tabular}
\label{tbl:parameters}
\end{table}
Matrix $\boldsymbol{D}$ contains the training data to design the detector via rank decomposition model. The rank decomposition model can be constructed via singular value decomposition (SVD) of $\boldsymbol{D}$ as follows, 
\begin{equation}
\small
\label{eq:SVD}
\boldsymbol{D}=\boldsymbol{U}\boldsymbol{\Lambda V}^T=\sum_{r=1}^{rank} \lambda_r \boldsymbol{u}_r\boldsymbol{v}_r^T,
\end{equation}
where, $\text{diag}{(\Lambda)}=[\lambda_i]$ contains singular values of $\boldsymbol{D}$. Transpose of $\boldsymbol{V}$ is indicated by  $\boldsymbol{V}^T$. Moreover, columns of $\boldsymbol{U}$ and $\boldsymbol{V}$ are right singular vectors and left singular vectors, respectively. The rank of $\boldsymbol{D}$ is upper bounded by $R_{\text{max}}$= min$(T,b)$. Rank of a matrix  is equal to the minimum number of rank-one components that holds Eq.~(\ref{eq:SVD}). However, the actual rank of a typical data could be much smaller than $R_{\text{max}}$  as the measurements from devices are dependent on each other. The goal of the outlier detection algorithm is to learn the possible dependencies and detect those patterns that do not agree with the rest of the measurements.  
By rejecting the inconsistent data, the blockchain network becomes AI-enabled. In AIBC, the devices associated with inconsistent data are  excluded to go to the next consensus step. The low-rank approximation of the measurements can be written as follows,
\vspace{-2mm}
\begin{equation}
\small
\label{eq:SVD-apprx}
\tilde{\boldsymbol{D}}=\sum_{r=1}^{R_{est}} \lambda_r \boldsymbol{u}_r\boldsymbol{v}_r^T,
\vspace{-2mm}
\end{equation}
where, $\tilde{\boldsymbol{D}}$ is the best low-rank approximation of $\boldsymbol{D}$.  Moreover, parameter $R_{est}$ can be estimated by analyzing the singular values. 
Rank is the minimum number of learned patterns such that data of all devices in a block can be represented  as a linear combination of those patterns. Using the training data and the estimated rank, a margin around the model can be identified. 
A straightforward criterion for rank estimation is based on the Frobenius norm of residual. 
\begin{equation}
\small
\label{eq:rank_est}
R_{est} =\text{min}\; R \;\;\text{\small{s.t.}}\;\; \frac{\|\boldsymbol{D}-\sum_{r=1}^R \lambda_r \boldsymbol{u}_r\boldsymbol{v}_r^T\|_F}{\|\boldsymbol{D}\|_F}\le \epsilon. 
\end{equation}
Here, $\epsilon$ is a constant between $0$ and $1$. As $\epsilon$ increases, the required rank decreases. Let us define matrix $\boldsymbol{Z}$ as the difference of $\boldsymbol{D}$ and $\tilde{\boldsymbol{D}}$, i.e., $\boldsymbol{Z}=\boldsymbol{D}-\tilde{\boldsymbol{D}}$. Each row of $\boldsymbol{Z}$ corresponds to the estimated perturbation of a device. The threshold for margin of each device is defined by $h_n$, which is a function of the desired false alarm probability of  detector.
The parameters of the employed low rank model are summarized in Table \ref{tbl:parameters}. The learned detector is characterized by $[ \boldsymbol{U},R_{est},\boldsymbol{h}]$. Vector $\boldsymbol{h}$ is concatenation of $[h_1, \cdots,\; h_d]$.


The input data at time slot $t$, $\boldsymbol{d}_t$, must be analyzed according to the learned detector by training data $\boldsymbol{D}$. First,   $\boldsymbol{d}_t$ should be projected on the span of the learned detector as:
\begin{equation}
\small
\label{eq:projection}
\tilde{\boldsymbol{d}}_t=\boldsymbol{U}(\boldsymbol{U}^T\boldsymbol{U})^{-1}\boldsymbol{U}^T \boldsymbol{d}_t,
\end{equation}
where, $\tilde{\boldsymbol{d}}_t$ is the projection of $\boldsymbol{d}_t$ on the low-rank model. The residual of the projection is defined as the difference of the measured data and the projected data on the model, i.e., $\boldsymbol{z}_t=\boldsymbol{d}_t-\tilde{\boldsymbol{d}}_t$. This residual vector contains the mismatch of $N$ devices for time $t$. The value of $z_t(n)$   is the metric for decision on detecting outliers, i.e., if $|z_t(n)|$ is greater than the threshold, $h_n$, the measured data violates the margin of the model. 







\begin{algorithm}[h!]
\small
\caption{Outlier rejection based on low-rank model}\label{alg:out_rej}
\algsetup{
linenosize=\small,
linenodelimiter=:
}
\begin{algorithmic}[1]
\REQUIRE training data ($\boldsymbol{D}$), online time series $\boldsymbol{d}_t$ and $P_{fa}$. 
\STATE compute $\boldsymbol{U}$, $\boldsymbol{\Sigma}$ and $\boldsymbol{V}$ using Eq. (\ref{eq:SVD}).\\
 \textbf{Initialization:}\\ 
\STATE model.rank $\leftarrow$ estimate using Eq. (\ref{eq:rank_est}).
\STATE model.bases $\leftarrow  \boldsymbol{U}(:,1:\text{model.rank})$. 
\STATE model.thresholds $\leftarrow$  select thresholds based on $P_{fa}$. \\
$ \text{for a new time slot collect}\; \boldsymbol{d}_t\;$ \text{from $n$ devices} 
\STATE $\tilde{\boldsymbol{d}}_t \leftarrow$ project $\boldsymbol{d}_t$ on the span of model Eq. (\ref{eq:projection}).
\STATE $\boldsymbol{z}=\boldsymbol{d}_t-\tilde{\boldsymbol{d}}_t$.
\\ $\qquad$ for each device ($n$) that $|z_n|>$model.thresholds($n$)
\STATE $\qquad \quad$ Outlier(t) $\leftarrow \;$ Outlier(t) $\bigcup\;\{n\}$ 
\\ $\qquad$ end for\\
\end{algorithmic}
\label{alg:out1}
\end{algorithm}

\vspace{-3mm}
\subsection{Performance Analysis}
\label{sec-performance}
A conventional hyperledger fabric network can tolerate up to $1/3$ malicious devices. However, proposed AIBC network can significantly increase this threshold by a carefully designed detector. There are two probabilities associated with an outlier detection algorithm: probability of detection ($P_{d}$) and probability of false alarm ($P_{fa}$). Probability of miss detection (complement of probability of detection) corresponds to the case that our algorithm fails to detect a malicious node. Probability of false alarm is associated with the scenario that the algorithm discerns an intact node as a malicious node. According to these two values, the fault tolerance of our proposed architecture can be improved in comparison with the naive PBFT consensus protocol. However, there exist the case that the performance of our architecture be less than $33.\overline{3}\%$ which is only possible if the detector is not well-designed such that it has a very high probability of false alarm or a very low probability of detection.
 
 The impact of the designed detector on the fault tolerance of the AIBC network is illustrated in the following inequality:
\begin{equation}
\label{eq:performance}
F_{det}=F_{raw}(1 - P_{d}) + (1 - F_{raw})P_{fa} \le \frac{1}{3}\;.
\end{equation}
In this inequality, $F_{det}$ denotes the  fault tolerance of the AIBC network over the filtered data using the designed detector. For successful execution of PBFT consensus in the second step consensus, $F_{det}$ should be less than $1/3$. $F_{raw}$ is the initial fault tolerance of our network in the first step before performing outlier detection which can be  greater than $1/3$. However, in a conventional hyperledger fabric, there is no detector and the  fault tolerance (equivalent to $F_{raw}$ in our AIBC network) should be less than $1/3$.
The effect of the detector on the threshold is investigated via implementation.

\section{Implementation Details and Results}
\label{sec-implemetation}
We implemented the proposed 3-layer AIBC network using hyperledger fabric framework version 1.1.0. The code is written in chaincode using {\tt Golang}. Following are the hardware specifications of the two laptops we used:  Core i7-6500U processor, CPU 2.5 GHz $\times$ 4, Ubuntu 18.04 LTS. 
First, we explain the architecture of the implemented network and the 
dataset we used. Then, we present the performance of the implemented AIBC network in terms of latency and the accuracy of the outlier detection algorithm.

\subsection{Network architecture and dataset}
We have simulated the application layer with $100$ sensors and devices using Matlab. These devices send their data to the AIBC network in each time unit. The  edge blockchain  and core blockchain layers are simulated on two separate but similar laptops. Different components of the AIBC network including all the peers and orderer are defined in separate containers using Docker. These containers can communicate with each other through a channel. The containers in the two laptops are connected using Docker swarm. 

The implemented chaincode has three main functions: \textit{init}, \textit{invoke}, and \textit{query}. Init function is used for initializing the number of sensors in layer 1 and their names, and initializing the blockchain with the first input data from the devices. Invoke function is used to receive new data from devices in layer 1. Each time an application sends new data to an aggregator, invoke function is executed which sends back the result after endorsement to that application. Query function is used to obtain the current value of a device in the IoT network. 

The detector is also implemented using {\tt Golang}. It has init and invoke functions. Init function initializes the peers with number of devices and their names and initializes primary data model in AIBC. The primary data model is obtained through outlier detection algorithm from a synthesized dataset. We use a $100 \times 100$ matrix where each column is new data obtained from an IoT network with $100$ devices. Invoke function is responsible for updating the data model through the outlier detection algorithm, detecting outlier data within a block, and discarding it.
After a block is sent to the peers by the orderer, invoke function of the detector is executed on each peer for execution of step-1 of the consensus protocol. It will discard the outlier data from that block and send the result to the peer to initiate the  step-2 of the consensus protocol.



\subsection{Network Latency}
There are several delays for the invoke function of the designed detector: outlier detection algorithm delay, model update delay, dataset update delay, and devices state update delay. Outlier detection algorithm delay is the time spent to execute this algorithm to infer outlier data in the latest $100$ readings from $100$ devices in the Layer $1$. Model update delay is the time it takes to update the model obtained by outlier detection algorithm. It should be mentioned that outlier detection  is a machine learning algorithm in which the learned detector is updated at each time based on the observed data.
The model is updated based on the last $100$ readings from different devices in Layer $1$. Dataset update delay is the delay associated with updating the last 100 readings from all $100$ devices in the ledger. Devices state update delay is the time spent to update new values of the devices in the ledger.
\begin{figure}[!t]
\begin{subfigure}[t]{0.23\textwidth}
        \centering
        \includegraphics[width=\textwidth, height=1.25in]{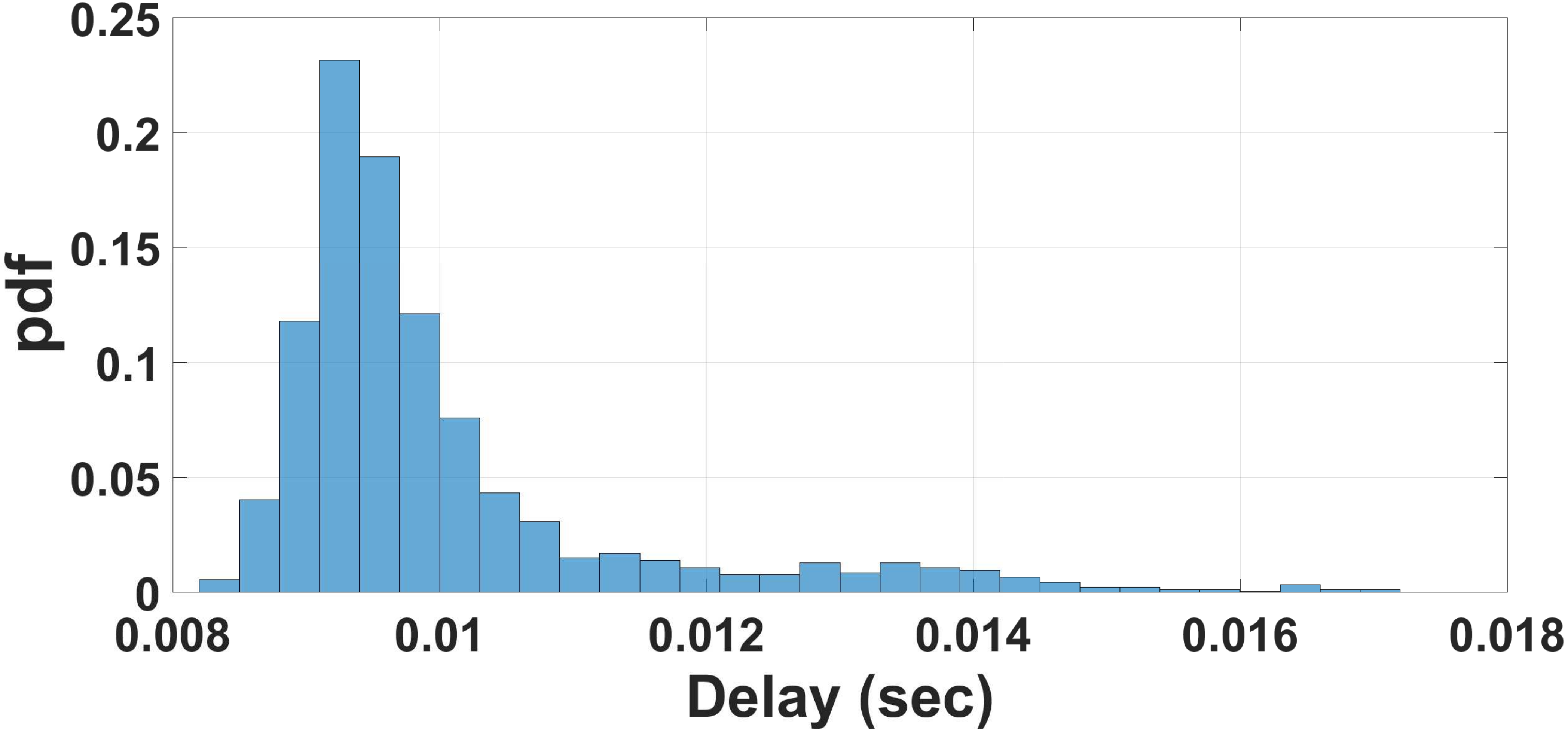}
        \label{latency2}
        \vspace{-4mm}
        \caption{}
    \end{subfigure}
    \begin{subfigure}[t]{0.23\textwidth}
        \centering
        \includegraphics[width=\textwidth, height=1.25in]{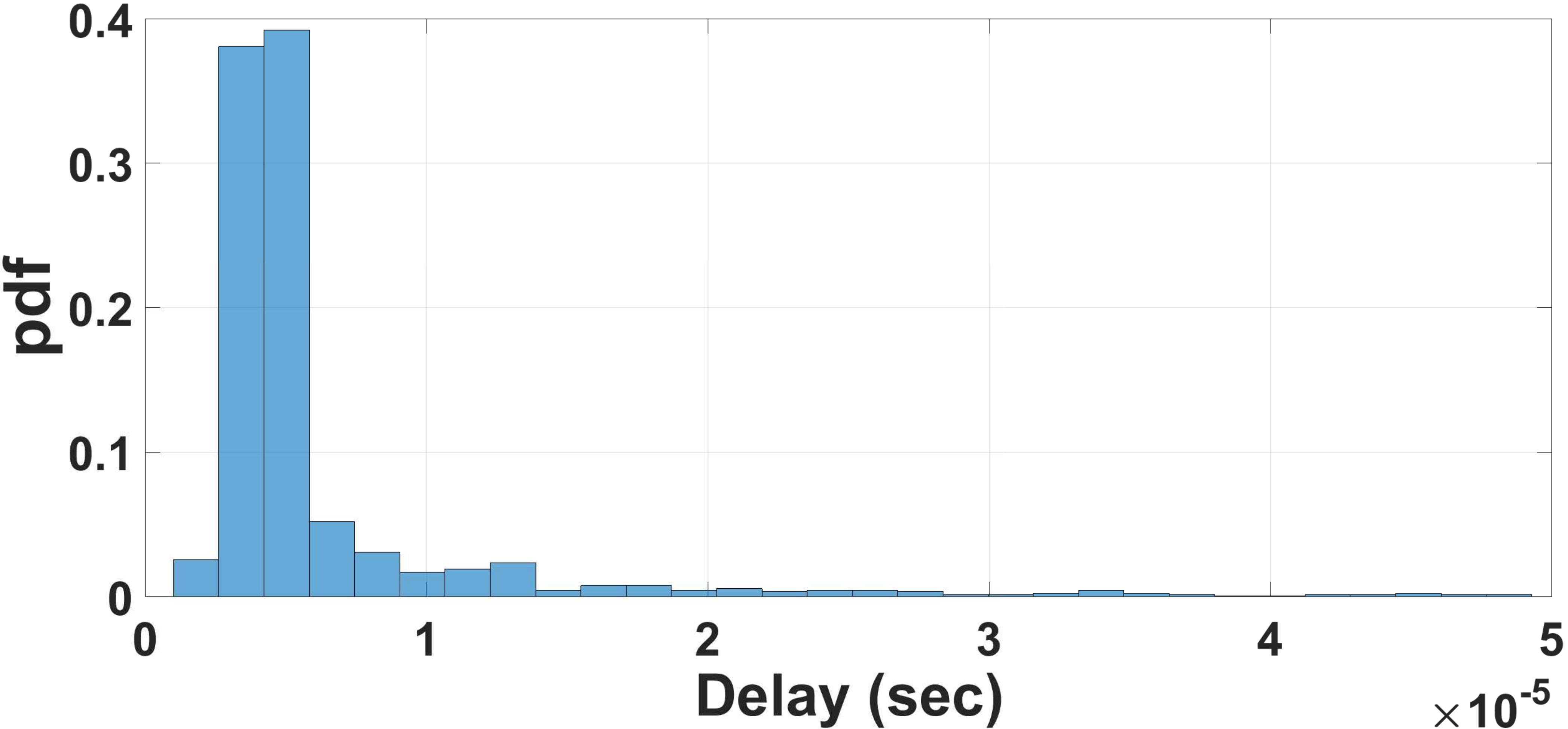}
         \label{outlierdevices}
         \vspace{-4mm}
        \caption{}
    \end{subfigure}
    \vspace{-2mm}
    \caption{\small{Acceptable and unavoidable delays. (a) Outlier detection algorithm delay, (b) Devices state update delay.}}
     \label{res:: delays1}
     \vspace{-3mm}
\end{figure}


\begin{figure}[!t]
\begin{subfigure}[t]{0.23\textwidth}
        \centering
        \includegraphics[width=\textwidth, height=1.15in]{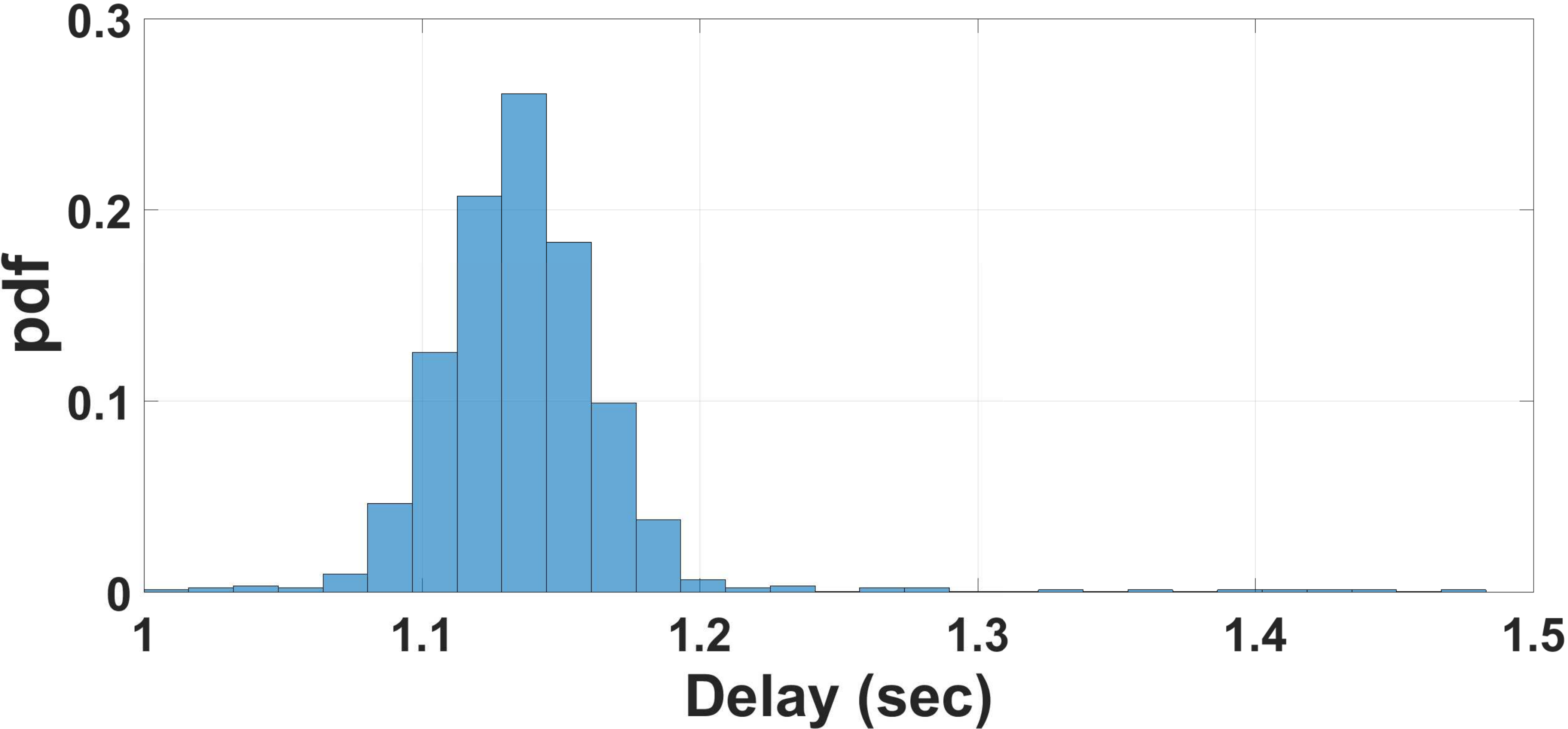}
        \label{latency2}
        \vspace{-5mm}
        \caption{}
    \end{subfigure}
    \begin{subfigure}[t]{0.23\textwidth}
        \centering
        \includegraphics[width=\textwidth, height=1.15in]{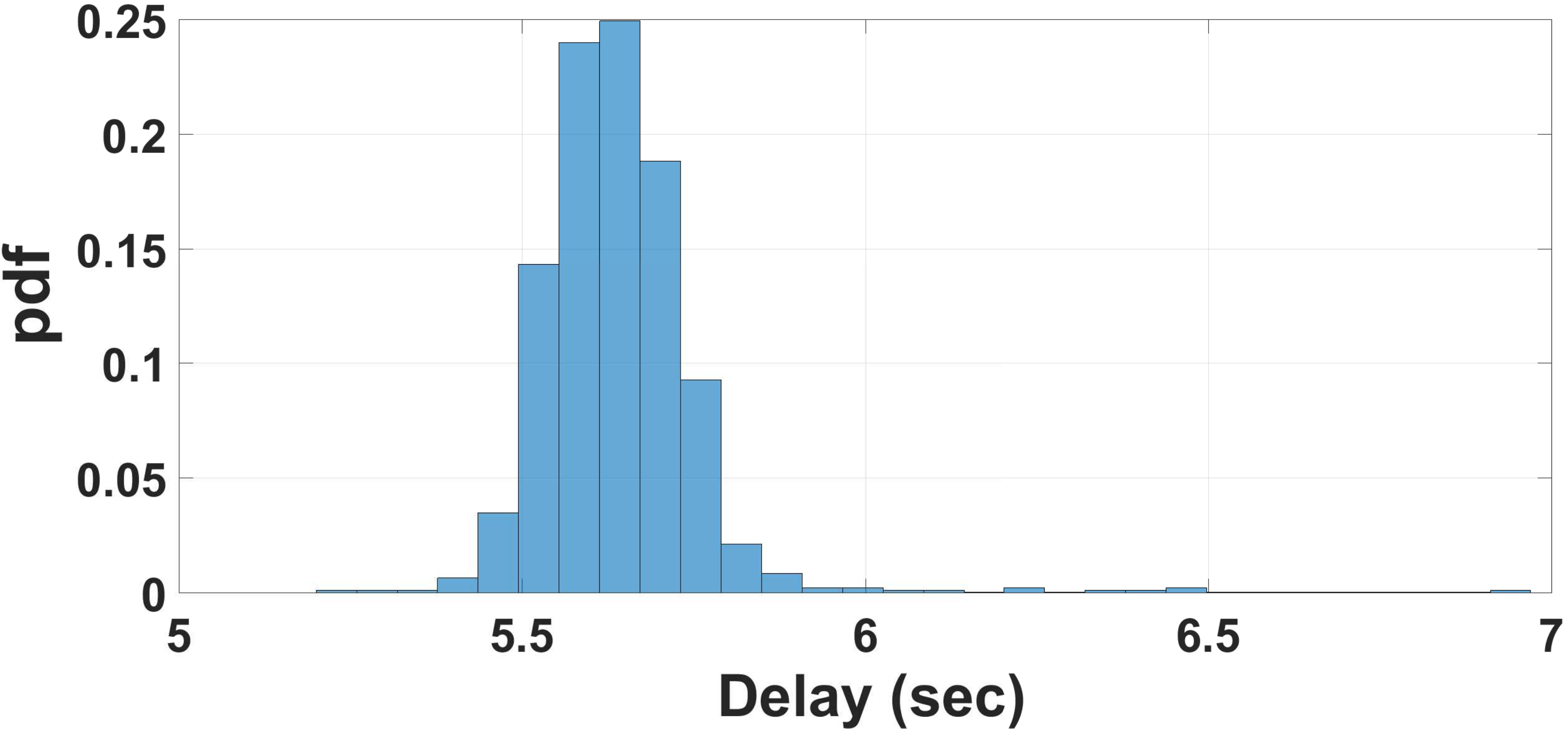}
         \label{outlierdevices}
         \vspace{-5mm}
        \caption{}
    \end{subfigure}
    \vspace{-2mm}
    \caption{\small{Unacceptable and avoidable delays. (a) Model update delay, (b) Dataset update delay.}}
     \label{res:: delays2}
     \vspace{-3mm}
\end{figure}

\begin{figure}[b]
\vspace{-5mm}
\centering
\includegraphics[width=2.75in,height=1.4 in]{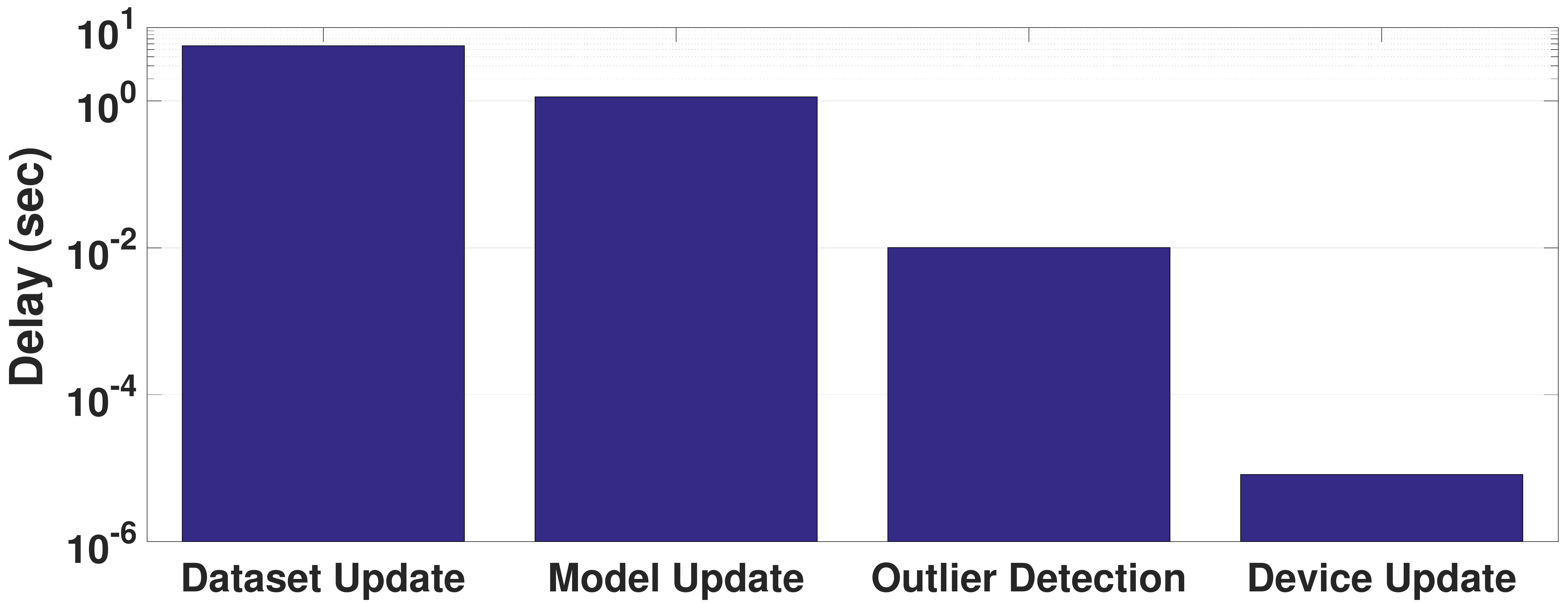}
\vspace{-2mm}
  \caption{\small{Latency of different sections of the AIBC network.}}
  \vspace{-3mm}
\label{latency2}
\end{figure}
We show the probability density function (pdf) of all the mentioned delays for our implemented IoT network with $100$ devices in $951$ time slots in Figs.~\ref{res:: delays1} and~\ref{res:: delays2}. A hyperledger fabric network reach consensus over a new block in milliseconds~\cite{salimitari2018overview} and adding that block to the copy of ledger would take about $5$ microseconds according to Fig.~\ref{res:: delays1}(b). These delays exist in any hyperledger fabric implementation and are acceptable for a smart home network. Our designed detector would add some new delay components for the network. Outlier detection algorithm would take about $9.5$ milliseconds as shown in Fig.~\ref{res:: delays1}(a) which will not affect the performance of a smart home network. However, according to Fig.~\ref{res:: delays2}, model update delay and dataset update delay will take about $1.14$ and $5.65$ seconds respectively which is not acceptable for a smart home network. However, dataset update delay and model update delay can be easily eliminated if we use pre-learned data using a proper dataset. Therefore, our proposed architecture will incur an additional delay of about $9.5$ milliseconds for outlier detection which is acceptable for a smart home network since consensus is reached in milliseconds. The delays for different sections of our implementation are compared in Fig.~\ref{latency2}.

\vspace{-1mm}
\subsection{Two-step Consensus Protocol Accuracy}

\begin{figure}[t]
\begin{subfigure}[t]{0.23\textwidth}
        \centering
        \includegraphics[width=\textwidth, height=1.55in]{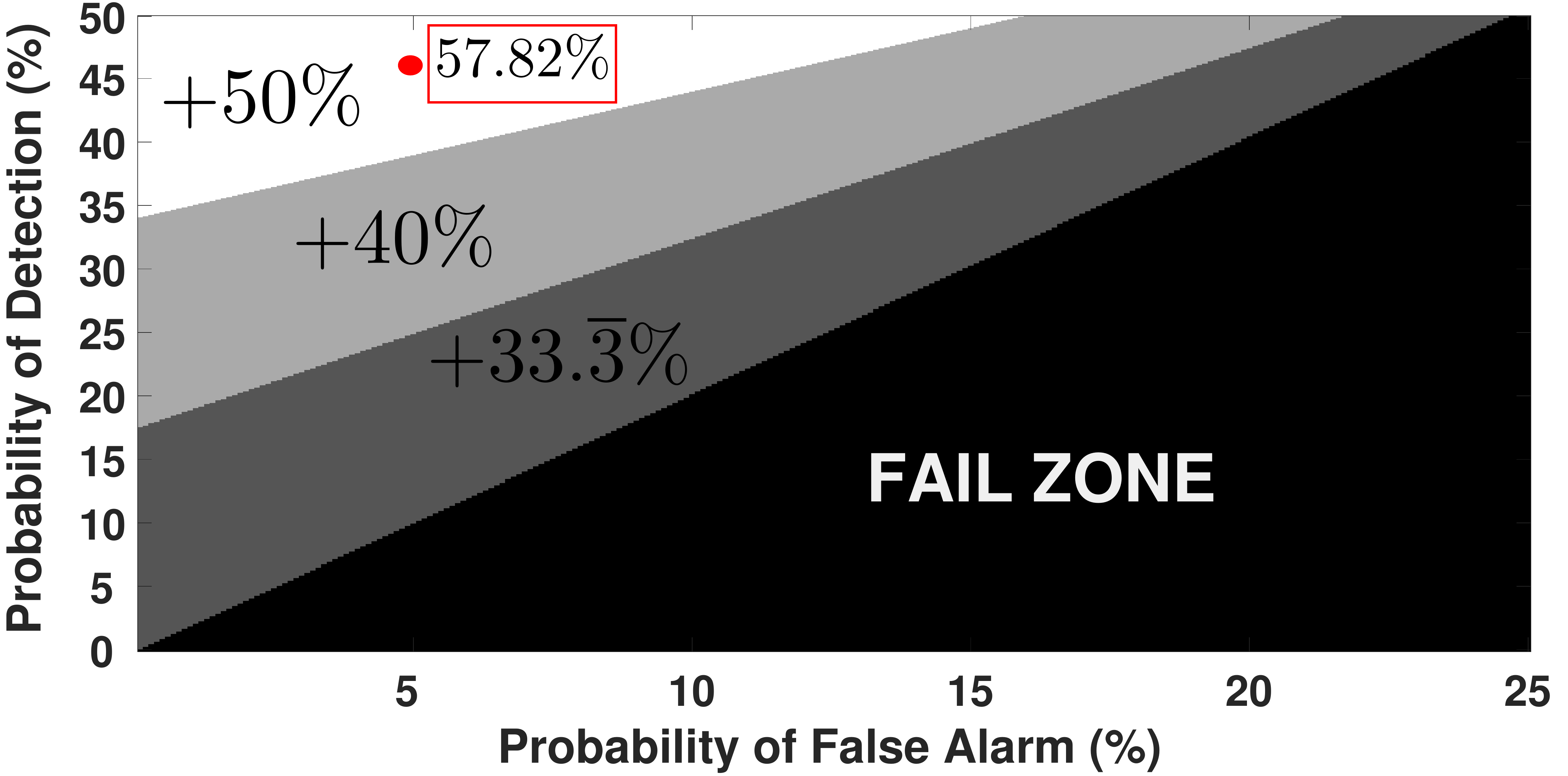}
        \label{performance}
        \vspace{-6mm}
        \caption{}
    \end{subfigure}
    \begin{subfigure}[t]{0.23\textwidth}
        \centering
        \includegraphics[width=\textwidth, height=1.55in]{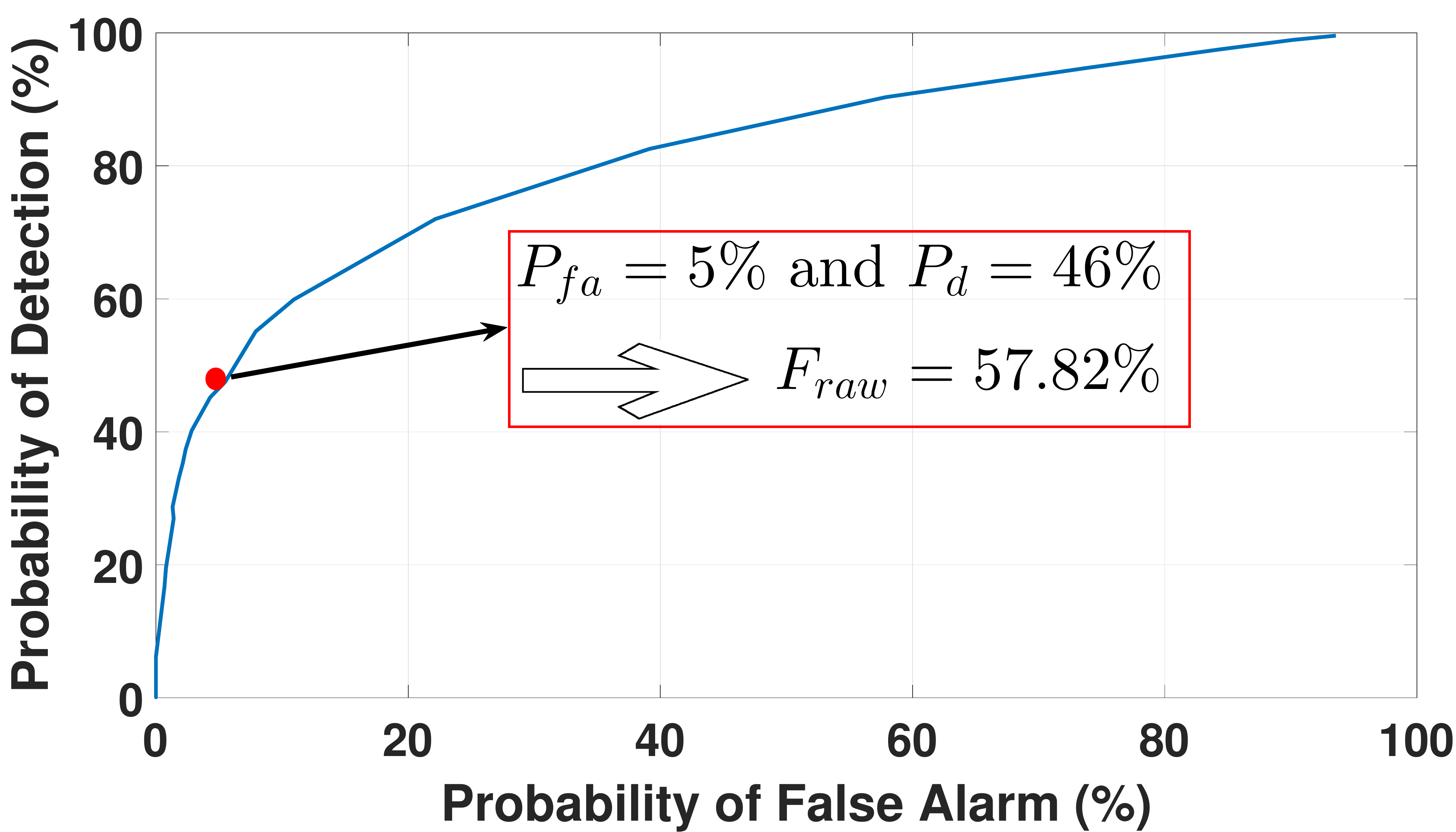}
         \label{error}
         \vspace{-6mm}
        \caption{}
    \end{subfigure}
    \vspace{-2mm}
    \caption{\small{Detector Performance. (a) Fault tolerance of the proposed algorithm, (b) Accuracy of the proposed outlier detector for a dataset with large number of faulty devices.}}
     \label{result}
     \vspace{-5mm}
\end{figure}

The last factor to evaluate our implementation is the accuracy of our proposed algorithm in terms of probability of detection and  probability of false alarm and its impact on the fault tolerance of the proposed architecture. 
The fault tolerance for different probabilities of detection and false alarm is illustrated in Fig.~\ref{result}(a). Fault tolerance of less than $33.\overline{3}\%$ is denoted as fail zone which is only possible if the detector is designed very inexpertly with a significantly high probability of false alarm or low probability of detection. 
In general, the performance of the network is enhanced and the fault tolerance of the AIBC network for different  detectors (different probability of detection and probability of false alarm) is found to be  more than $33.\overline{3}\%$, more than $40\%$, or more than $50\%$ as shown in Fig.~\ref{result}(a). 


Fig.~\ref{result}(b) shows the accuracy of our algorithm on a  synthesized dataset with large number of faulty devices. Although there exists a large number of malicious devices (outlier data) in the dataset, the detector is learned such that the fault tolerance of the network is more than $50\%$. This is inferred by choosing any operating point in Fig.~\ref{result}(b) and the corresponding point in Fig.~\ref{result}(a). As an example, an operating point of the detector is shown at $P_{fa}=5\%$ and $P_d=46\%$. The fault tolerance of the network in this point is $57.82\%$ according to Inequality (\ref{eq:performance}).


\vspace{+1mm}
\section{Conclusions}
We proposed the AIBC network with a 2-step consensus protocol using outlier detection algorithm and PBFT. 
Outlier detection algorithm acts as the first step consensus and verifies the compatibility of new data and discards the suspicious ones in order to increase fault tolerance of the network for the second step consensus (PBFT). We measured the latency, accuracy, and performance of our method. 
Results reveal significant increase in the fault tolerance of hyperledger fabric by our detector. We employed an outlier detection scheme, however, recent advances in artificial intelligence such as deep learning and reinforcement learning can be exploited for designing the detector in a more robust manner. 
 \vspace{+1mm}

\end{document}